\documentstyle[aps,prl,multicol,epsf]{revtex}
\begin{document}
\author{X. X. Yi$^{1}$\footnote{E-mail:yixx050@nenu.edu.cn} H. T. Cui$^1$ and  X.G.Wang$^2$}
\address{$^1$ Institute of Theoretical Physics, Northeast Normal University, Changchun
130024, China\\
$^2$ Department of Physics, Macquarie University, Sydney, New
South Wales 2109, Austrilia}
\title{Dynamics of the entanglement rate in the presence of decoherence}
\maketitle
\begin{abstract}
The dynamics of the entanglement rate  are investigated  in this
paper for pairwise interaction and two special sets of initial
states. The results show that for the  given interaction and the
decoherence scheme, the competitions between decohering and
entangling lead to two different results--some initial states  may
be used to prepare entanglement while the others do not. A
criterion on decohering and entangling
is also presented and discussed.\\
{\bf PACS number(s):03.67.-a, 03.65.Bz,03.67.Hk,03.65.Ca}
\end{abstract}
\vspace{4mm}
\begin{multicols}{2}[]
Entanglement plays an essential role in quantum information
theory, the sharing of entanglement between sender and receiver
allows for quantum teleportation[1], quantum superdense coding[2]
and  the other applications to quantum information processing[3].
Creating entanglement in a proper way is thus an important issue.

In general, entanglement between two systems can be generated if
they interact in a controlled way. However,  for a practical
experiment, the production of entanglement is very difficult due
to the weak interaction between the systems. Thus how to improve
efficiency of the production by using those interactions become a
very relevant problem. Very recently, D\"ur {\it et al.} [4]
consider a situation that one has a given non-local Hamiltonian
and ask, what is the most efficient way of entangling particles?
Their answers are that (i) the initially entangled two particles
can improve the efficiency of the production, and (ii) one can
also improve the efficiency by using some ancillas.

In this paper, we shed light on this issue again by  taking the
decoherence effects into account. As you will see, the problem for
mixed states is complicated, thus we choose two special sets of
mixed state to study the problem. This is the limitation of this
paper. The results  show that there is a competition between
entangling and decohering in the entanglement production process,
some initial states may work very well in the absence of
decoherence, but they do not provide the best way to produce
entanglement in the presence of decoherence.

 To begin with, we
recall the definition of entanglement rate $\Gamma(t)$[4]
\begin{equation}
\Gamma(t)=\frac{d E(t)}{dt},
\end{equation}
where $E(t)$ denotes an entanglement measure of a state $\rho(t)$.
In this paper, we pay our attention first to the case of two
qubits, and then generalize the discussion to the case of d-level
system  with $d>2$. We use the following notations throughout this
paper: $|\xi_i\rangle (i=1,2,3,4)$ stand for bases of the
two-qubit system, $|\xi_1\rangle=|00\rangle$,
$|\xi_2\rangle=|01\rangle$,$|\xi_3\rangle=|10\rangle$,
$|\xi_4\rangle=|11\rangle$. $|m\rangle (m=0,1)$ denotes the two
states of one qubit, and
$\rho_{ij}=\langle\xi_i|\rho(t)|\xi_j\rangle$ represent the matrix
elements of $\rho(t)$ in the space spanned by $|\xi_i\rangle
(i=1,...,4)$. With those notations, we choose 15 independent
variables to describe a general two-qubit state $\rho(t)$, they
consist of 3 independent diagonal elements
$\rho_{11},\rho_{22},\rho_{33}$  and 6 complex off-diagonal
elements $\rho_{ij}(i,j=1,...,4,$ and $j>i)$ of matrix
$\rho(t)$[5].

In order to calculate the entanglement rate, we have to express
the entanglement measure of the state $\rho(t)$ as a function of
$\rho_{ij}(i,j=1,...,4, j\geq i)$. For an entangled two-qubit
system, we may choose the Wootters concurrence as the entanglement
measure

\begin{equation}
E(t)={\cal E}(c(\rho)),
\end{equation}
with
$$ {\cal E} (c)=h(\frac{1+\sqrt{1-c^2}}{2}),$$
$$ h(x)=-xlog_2 x-(1-x)log_2(1-x),$$
$$ c(\rho)=max\{0,\lambda_1-\lambda_2-\lambda_3-\lambda_4\},$$
where the $\lambda_i$s are the square roots of the eigenvalues of
the non-Hermitian matrix $\rho\tilde{\rho}$ with
$\tilde{\rho}=(\sigma_y\otimes
\sigma_y)\rho^*(\sigma_y\otimes\sigma_y)$ in decreasing order. In
the  space spanned by $\{\xi_i\rangle, i=1,2,3,4\},$
$\tilde{\rho}$ reads

\begin{equation}
\tilde{\rho}= \left(\matrix{ \rho_{44}
&-\rho_{34}&-\rho_{24}&\rho_{14}\cr
-(\rho_{34})^*&\rho_{33}&\rho_{23}&-\rho_{13}\cr
-(\rho_{24})^*&(\rho_{23})^*&\rho_{22}&-\rho_{12}\cr
(\rho_{14})^*&-(\rho_{13})^*&-(\rho_{12})^*&\rho_{11} } \right ).
\end{equation}
 The Wootters concurrence gives an
explicit expression for the entanglement of formation, which
quantifies the resources needed to create a given entangled state.
Note that the Wootters concurrence is a function of
$\rho_{ij}(i,j=1,...,4,j\geq i)$. Therefore, we can write
\begin{equation}
\Gamma(t)=\sum_{i,j=1,j\geq i}^4\frac{\partial E}{\partial
\rho_{ij}}\frac{\partial \rho_{ij}}{\partial t},(\rho_{ij}\neq
\rho_{44}).
\end{equation}
Eq.(4) shows that given a particular entanglement measure
$E(\rho_{ij})$, we just have to determine $\partial
\rho_{ij}/\partial t$ . In order to do that, we need to find the
time evolution of the state $\rho(t)$ in the presence of an
environment. Generally speaking, interactions between a quantum
system and its environment result in two kinds of irreversible
effects: dissipation and dephasing. The first effect is due to the
energy exchange between the system and its environment, whereas
the second one comes from the system-environment interaction that
does not change the system energy. Both  dissipation and dephasing
lead to decoherence. In what follows, we first drive an expression
for the time derivative of $\rho(t)$ in terms of Kraus operators,
this equation may be useful in the case of the Krause operators
are easily given. Then we adapt the other description of
decoherence to put forward our discussion. Consider a quantum
system of two qubits $\rho$ interacting with an environment
$\rho_E=\sum_{\nu}p_{\nu}|\nu\rangle\langle \nu|$, after a finite
time evolution governed by unitary evolution operator $U(t,0)$,
the total density operator(the system plus the environment)
$\rho_t(t)$ is given as
\begin{equation}
\rho_t(t)=U(t,0)(\rho_E\otimes\rho)U^{\dagger}(t,0).
\end{equation}
Taking a partial trace over environment variables we can get the
density operator of the two-qubit system in  the following form[6]
\begin{equation} \rho(t)=\mbox{Tr}_E\rho_t(t)=\sum_{\mu,\nu}
K_{\mu\nu}(t,0)\rho(0)K_{\mu\nu}^{\dagger}(t,0),
\end{equation}
where $K_{\mu\nu}(t,0)=\langle
\mu|\sqrt{p_{\nu}}U(t,0)|\nu\rangle.$ The Kraus operators
$K_{\mu\nu}$ satisfy $\sum_{\mu\nu} K_{\mu\nu}^{\dagger}
K_{\mu\nu}=1.$ No environment around the two-qubit system
indicates that there is only one term in the sum eq.(6). In weak
system-environment interaction limit, the density operator of the
environment $\rho_E$ remains unchanged in the whole time evolution
process, this approximation can be found in the derivation of the
master equation, which we will discuss later on. Under weak
system-environment interaction, the Kraus operators can be
expanded to first order of $dt$ as
\begin{eqnarray}
K_{\mu\nu}(t+dt,t)&=&\langle
\mu|\sqrt{p_{\nu}}U(t+dt,t)|\nu\rangle\nonumber\\
&\simeq&\sqrt{p_{\nu}}\delta_{\mu\nu}-idt
\sqrt{p_{\nu}}\langle\mu|H_t|\nu\rangle,
\end{eqnarray}
where $H_t$ stands for the Hamiltonian of the total
system(two-qubit plus its environment), and this equation holds
only for very small $dt$. Substituting Eq.(7) into Eq.(6), we
obtain $\rho(t+dt)$ in terms of $\rho(t)$(to first order of $dt$)
\begin{eqnarray}
\rho(t+dt)&=&\rho(t)-idt\sum_{\mu,\nu}\sqrt{p_{\nu}}\langle\mu|H_t|\nu\rangle\rho(t)\nonumber\\
&+&idt\rho(t)\sum_{\mu,\nu}\sqrt{p_{\nu}}\langle\mu|H_t|\nu\rangle\nonumber\\
&\equiv&\rho(t)+d\rho(t).
\end{eqnarray}
In order to calculate the entanglement rate, we have to compute
$\partial\rho_{ij}/\partial t$. Using standard perturbation
theory, we find $\partial\rho_{ij}/\partial t$ as follows
\begin{eqnarray}
 \frac{\partial\rho_{ij}}{\partial t}=\langle\xi_i|\partial\rho(t)/\partial t|\xi_j\rangle
 \end{eqnarray}
with
$$
\frac{\partial \rho(t)}{\partial
t}=-i[\sum_{\mu,\nu}\sqrt{p_{\nu}}\langle\mu|H_t|\nu\rangle,\rho(t)]$$
Eq.(8) and Eq.(9) show that
$\sum_{\mu,\nu}\sqrt{p_{\nu}}\langle\mu|H_t|\nu\rangle$ play a
role of effective Hamiltonian for the two-qubit system, in this
sense we may rewrite Eq.(8) in the following form
\begin{equation}
i\frac{\partial \rho}{\partial t}=[H_e,\rho],
\end{equation}
where $H_e=\sum_{\mu,\nu}\sqrt{p_{\nu}}\langle\mu|H_t|\nu\rangle.$
This expression is useful and easy to handle when we know the
Kraus operators.

The other tool to study the quantum dissipative system is the
master equation, which can be obtained in Markovian limit [7-9].
This approximation is  very useful because it is valid for many
physical relevant situations and  its numerical solutions can be
easily found. As given by Gardiner, Walls and Millburn, Louisell
in their textbook [9], the reduced density matrix $\rho$ of the
open system which is linearly coupled to its environment obeys the
following master equation of Lindblad form [10]
\begin{eqnarray}
\dot{\rho}(t)&=&-i[H_0,\rho]\cr &+&\frac 1 2 \sum_m K_m(2X_m^-\rho
X^+_m- X_m^+X_m^-\rho-\rho X_m^+X_m^-)\nonumber\\ &+&\frac 1 2
\sum_mG_m(2X_m^+\rho X^-_m- X_m^-X_m^+\rho-\rho X_m^-X_m^+)
\end{eqnarray}
with $$K_m=2 \mbox{Re}\left[\int_0^{\infty}d \tau e^{i\omega_m
\tau}{\rm Tr}_{env}\{A_m(\tau) A^{\dag}_m(0)\rho_{env}\}\right],$$
$$G_m=2 \mbox{Re}\left[\int_0^{\infty}d \tau e^{i\omega_m
\tau}{\rm Tr}_{env}\{A_m^{\dag}(\tau) A_m(0)\rho_{env}\}\right].$$
Here, $\rho(t)=\rho(t,K_m,G_m)$ stands for the density operator of
the system and $\rho_{env}$ denotes the density operator of the
environment, $X_m^{\pm}$ are eigenoperators of the system
satisfying $[H_0,X^{\pm}_m]=\pm\hbar\omega_m X_m^{\pm}$,
 $H_0$ stands for the free Hamiltonian of the system, and
 $A_m$($A_m^{\dagger}$) are operators of the environment through
 which the system and its environment couples together.
 Notice from Eq.(11) that $G_m$ should vanish at zero temperature
$T=0$, while $K_m$ should not if $A_m$ are indeed destruction
operators of some kind.

The time derivative of the matrix elements $\rho_{ij}$ in this
case is
\begin{equation}
\frac{\partial \rho_{ij}}{\partial t} =\frac{\partial
\rho_{ji}^*}{\partial t}=\langle \xi_i|\dot{\rho}|\xi_j\rangle.
\end{equation}

Further more, we consider a case of a two-qubit system coupling to
 environments that consists of a set of harmonic oscillators. In
this case, the master equation takes the following form at zero
temperature
\begin{eqnarray}
\dot{\rho}(t)&=&-i[H_s\rho-\rho
H_s]\nonumber\\
&+&\frac{\gamma}{2}\sum_{i=1}^2(2\sigma_-^i\rho\sigma_+^i-\sigma_+^i\sigma_-^i\rho-\rho
\sigma_+^i\sigma_-^i),
\end{eqnarray}
where $H_s$ is the system Hamiltonian, which governs time
evolution of the two-qubit system in the absence of its
environment, $\sigma_+^i(\sigma_-^i)$ are pauli matrices, and
$\gamma$ represents the damping rate. Considering a system
Hamiltonian
\begin{equation} H_s=
\frac{\hbar\omega\sigma_z^1}{2}+\frac{\hbar\omega\sigma_z^2}{2}+\hbar
g(\sigma_+^1\sigma_-^2+\sigma_-^1\sigma_+^2)
\end{equation}
and substituting Eq.(13) into Eq.(12), we obtain(setting
$\hbar=1$)
\begin{eqnarray}
\dot{\rho_{11}}&=&\gamma(\rho_{22}+\rho_{33}),\nonumber\\
\dot{\rho_{22}}&=&-ig\rho_{32}+ig\rho_{23}+\gamma\rho_{44}-\gamma\rho_{22},\nonumber\\
\dot{\rho_{33}}&=&-ig\rho_{23}+ig\rho_{32}+\gamma\rho_{44}-\gamma\rho_{33},\nonumber\\
\dot{\rho_{44}}&=&-2\gamma\rho_{44},\nonumber\\
\dot{\rho_{12}}&=&ig\rho_{13}+\gamma\rho_{34}-0.5\gamma\rho_{12}+i\omega\rho_{12},\nonumber\\
\dot{\rho_{13}}&=&ig\rho_{12}+\gamma\rho_{24}-0.5\gamma\rho_{13}+i\omega\rho_{13},\nonumber\\
\dot{\rho_{14}}&=&-\gamma \rho_{14}+2i\omega\rho_{14},\nonumber\\
\dot{\rho_{23}}&=&-ig\rho_{33}+ig\rho_{22}-\gamma\rho_{23},\nonumber\\
\dot{\rho_{24}}&=&-ig\rho_{34}-1.5\gamma\rho_{24}+i\omega\rho_{24},\nonumber\\
\dot{\rho_{34}}&=&-ig\rho_{24}-1.5\gamma
\rho_{34}+i\omega\rho_{34}
\end{eqnarray}
The Hamiltonian $H_s$(14) describes two two-level atoms with
dipole-dipole interactions, which are a source of creating
entanglement for trapped atoms in an optical lattice[11,12]. For
an initial state, if the interaction terms(with coupling constant
$g$ in Eq.(15)) have no effects in the time evolution process,
it(an example is given below) could not be used to create
entangled state or to increase entanglement. Initial state with
the form of
\begin{equation}
\rho_0= \left(\matrix{ \frac{c+d}{2}&0 &0&\frac{d-c}{2}\cr
0&\frac{a+b}{2}&\frac{b-a}{2}&0\cr
0&\frac{b-a}{2}&\frac{a+b}{2}&0\cr \frac{d-c}{2}&0&0
&\frac{c+d}{2}}\right ).
\end{equation}
is a family of such initial states. In terms of Bell bases,
eq.(16) can be written as
$$\rho_0=a|\psi^+\rangle\langle \psi^+|+b|\psi^-\rangle\langle\psi^-|+
c|\phi^+\rangle\langle\phi^+|+d|\phi^-\rangle\langle\phi^-|.
$$
This is just the Werner state and any two-qubit entangled state
can be expressed in this form by performing a random bilateral
rotation on each shared pair[3]. We may use this kind of entangled
state to demonstrate how it decoheres, although it could not be
used to increase entanglement. We would like to mention that for a
general mixed state characterized by 15 independent parameters,
the problem becomes complicated. Hence we choose two special sets
of initial states to get some insights into the formalism.  In
terms of Wootters concurrence, the entanglement measure for
states(16) is
\begin{equation}
E(t)=-xlog_2x-(1-x)log_2(1-x),
\end{equation}
where $x=(1+\sqrt{1-F^2})/2$, $ F=a-b-c-d$ for $a-b-c-d>0$, or 0
for $a-b-c-d<0$, and we assume $a>b, a>c, a>d$ without loss of
generality.

It is easy to show that the entanglement rate defined by (1) takes
the following form($a+b+c+d=1$)
\begin{equation}
\Gamma(t)=\frac{\partial E(t)}{\partial a}\frac{\partial
a}{\partial t},
\end{equation}
with \begin{eqnarray} \frac{\partial E(t)}{\partial
a}&=&(log_2x-log_2(1-x))\frac{F}{\sqrt{1-F^2}},\nonumber\\
\frac{\partial a}{\partial t}&=&\frac{c+d}{2}\gamma -a\gamma.
\end{eqnarray}
By definitions, $F>0$ and $x>\frac 1 2$, then $\frac{\partial
E}{\partial a}>0$. So, wether the entanglement increase or
decrease for the initial states (16) only depends on
$\frac{\partial a}{\partial t}$. And $\frac{\partial a}{\partial
t}$ is always below zero. Therefore, we could not increase
entanglement starting from the initial states (16). Because
$\frac{\partial E}{\partial a}$ does not depend on $c$ and $d$,
the entanglement rate depends on $c$ and $d$ linearly, and with
$c+d$ increase, the entanglement decreases. Whereas the
entanglement rate is inversely  proportional to $\gamma$. The
dependence of $\Gamma(t)$ on parameter $a$ is shown in figure 1.

\begin{figure}
\epsfxsize=9cm \centerline{\epsffile{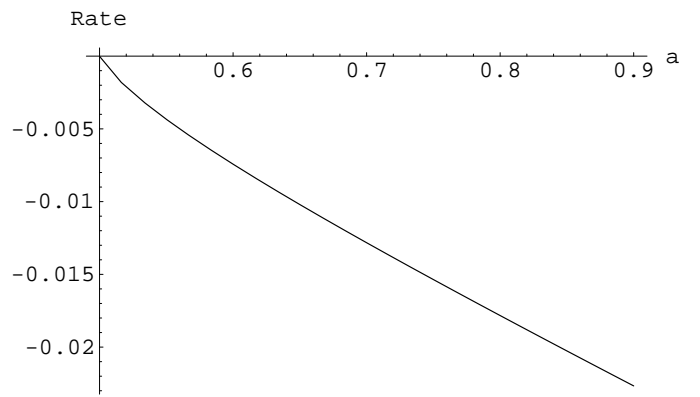}} \vskip 0.2cm
\caption[]{ Dependence of the entanglement rate $\Gamma$ on $a$
with $\gamma=0.01$ and $c+d=0.1$.}
\end{figure}
This figure shows that the larger the parameter $a$, the smaller
the entanglement rate. For a limit case $a=1$, and $c=d=b=0$ that
corresponds to the maximally entangled state $|\psi^+\rangle$, the
entanglement rate takes its minimum over states (16).

 In contrast to example eq.(16), we present here another kind of
 states
\begin{equation}
\rho_0= \left(\matrix{ 0 &0 &0&0\cr 0&p &q&0\cr 0&q^*&1-p&0\cr
0&0&0&0}\right ).
\end{equation}
This kind of state is of interest because entanglement contained
in those states range from zero to one(maximally entangled state).
And it is a typical family of states for a ring of $N$ qubits in a
translation  invariant quantum state[13].

The entanglement measure for this family of states has the same
expression as Eq.(17) except for replacing $F$ by $G$
\begin{eqnarray}
G=2|q|.
\end{eqnarray}
The positivity of the state $\rho_0$ Eq.(20) require
\begin{equation}
R=p^2-p+|q|^2\leq 0,
\end{equation}
this indicates that $|q|\leq \frac 1 2 $ for all family of states
(20). And for $|q|=\frac 1 2 $, there is only one value available
for $p$, i.e. $p=\frac 1 2 $. This is shown in  Fig.2 which gives
the dependence of $R$ on $p$ and $|q|$.

\begin{figure}
\epsfxsize=9cm \centerline{\epsffile{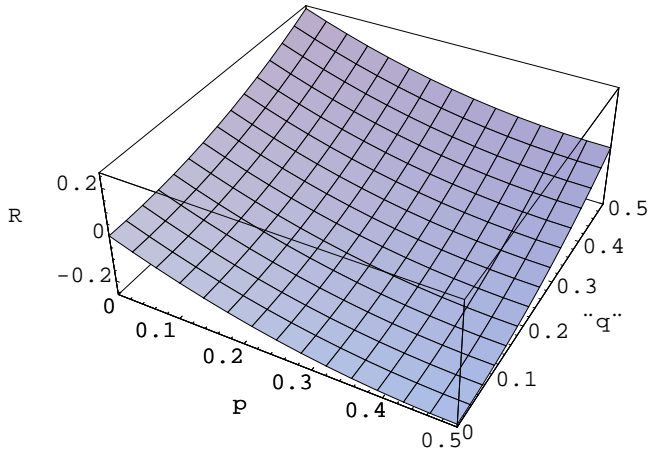}} \vskip 0.2cm
\caption[]{ Dependence of the quantity $R$ on $p$ and $|q|$, that
shows which region of $|q|$ and $p$ are available for the
state(20).}
\end{figure}

The entanglement rate in this example is
\begin{equation}
\Gamma(t)=\frac{\partial E(t)}{\partial p}\frac{\partial
p}{\partial t}+\frac{\partial E(t)}{\partial q^R}\frac{\partial
q^R}{\partial t}+\frac{\partial E(t)}{\partial q^I}\frac{\partial
q^I}{\partial t},
\end{equation}
where $q^{R(I)}$ represents the real (imaginary) part of $q$.
Eq.(15) and (17) together give
\begin{eqnarray}
\frac{\partial E(t)}{\partial
p}&=&0,\nonumber\\
\frac{\partial p}{\partial t}&=&-2gq^I+\frac 1 2 \gamma
(1-2p),\nonumber\\
\frac{\partial E(t)}{\partial
q^I}&=&(log_2 y-log_2(1-y))\frac{G}{\sqrt{1-G^2}}\frac{\partial G}{\partial q^I},\nonumber\\
\frac{\partial q^I}{\partial t}&=&g(2p-1)-\gamma q^I,\nonumber\\
\frac{\partial E(t)}{\partial
q^R}&=&(log_2y-log_2(1-y))\frac{G}{\sqrt{1-G^2}}\frac{\partial G}{\partial q^R},\nonumber\\
\frac{\partial q^R}{\partial t}&=&- \gamma q^R,\nonumber\\
\frac{\partial G}{\partial q^{R(I)}}&=&\frac{4q^{R(I)}}{|q|},
\end{eqnarray}
where $y=\frac 1 2 (1+\sqrt{1-G^2})$. Equations (23) (24) together
give
$$\Gamma(t)=log_2\frac{y}{1-y}\frac{G}{\sqrt{1-G^2}}[\frac{4gq^I(2p-1)-4\gamma|q|^2}{|q|}]  $$
This shows that there are competitions between entangling and
decohering. If $g/\gamma>|q|^2/(q^I(2p-1))$, $\Gamma(t)>0$,
entanglement increases. Otherwise the entanglement decreases. In
other words, in order to get a positive entanglement rate, the
decoherence rate $\gamma$ and the coupling constant $g$ should
satisfy condition $g>\gamma |q|^2/(q^I(2p-1))$
 For parameters $p=0.6, g=0.2,\gamma=0.01 $, the
entanglement rate $\Gamma$ versus $q^I$ and $q^R$ is illustrated
in Fig.3. \vspace{0.5cm}
\begin{figure}
\epsfxsize=9cm \centerline{\epsffile{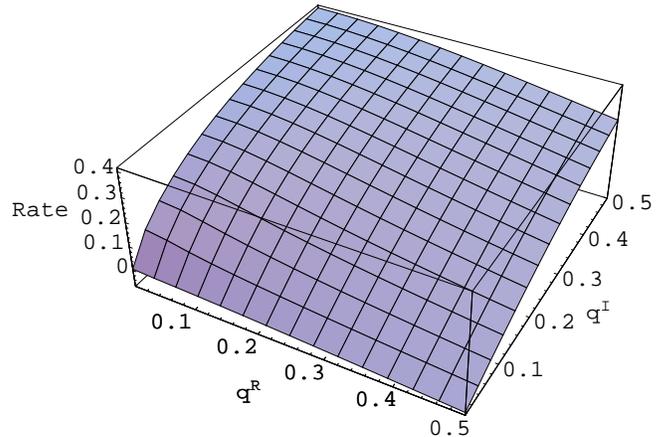}} \vskip 0cm
\caption[]{ Entanglement rate for the family of state (20) versus
$q^I$ and $q^R$. The parameters chosen are $p=0.6, g=0.2,
\gamma=0.01.$ }
\end{figure}

 The maximum of the entanglement rate is 0.4
corresponding $q^I=0.5, q^R=0.0$, while the minimum of $\Gamma$ is
-0.1 at about $ q^I=0$, $q^R=0.5$. Similarly, the maximum and
minimum of the entanglement rate change with $g$ and $\gamma$, but
$q^I_{max(min)}, q^R_{max(min)}$ which corresponding to
$\Gamma_{max}(\Gamma_{min})$ do not.

 In the end of this paper, we generalize the
formulas derived above to the case of multilevel systems, we
denote by $\sigma_i$ the generators of the group $SU(N)$ with $N$
being the dimension of the Hilbert space of the system $A$, and
$\tau_i$ the generators corresponding to system $B$ with dimension
$M$. With this notation, we may write any density matrix of the
composite system $A$ and $B$ in the following general form
\begin{equation}
\rho_{AB}=\frac{1}{MN}(1+\sum_i\alpha_i\sigma_i+\sum_j\beta_j\tau_j+\sum_{ij}\gamma_{ij}
\sigma_i\otimes\tau_j).
\end{equation}
Here we choose $\alpha_i$, $\beta_i$ and $\gamma_{ij}$ as the
independent variables to characterize the state of system $A$ plus
$B$.  The entanglement measure $E(\alpha_i, \beta_j,\gamma_{ij})$
is thus a function of $\alpha_i$, $\beta_j$, and $\gamma_{ij}$. It
is natural to express the entanglement rate in the following way
\begin{equation}
\Gamma(t)=\sum_i\frac{\partial E}{\partial
\alpha_i}\frac{\partial\alpha_i}{\partial t}+\sum_i\frac{\partial
E}{\partial \beta_i}\frac{\partial\beta_i}{\partial
t}+\sum_{ij}\frac{\partial E}{\partial
\gamma_{ij}}\frac{\partial\beta_{ij}}{\partial t}.
\end{equation}
Given an entanglement measure $E(\alpha_i,\beta_j,\gamma_{ij})$,
the derivatives $\partial E/\partial \alpha_i$, $\partial
E/\partial \beta_j$ and $\partial E/\partial \gamma_{ij}$ can be
easily calculated. The remained task is only to determine the time
derivative of $\alpha_i$, $\beta_j$, and $\gamma_{ij}$. Proceeding
as before, we obtain
\begin{eqnarray}
\frac{\partial \alpha_i}{\partial t}&=& {\mbox
Tr_A}(\sigma_i\frac{\partial \rho_{AB}(t)}{\partial t}),\, \, \,
\frac{\partial \beta_i}{\partial t}= {\mbox
Tr_B}(\tau_i\frac{\partial \rho_{AB}(t)}{\partial
t}),\nonumber\\
\frac{\partial \gamma_{ij}}{\partial t}&=&{\mbox
Tr_{AB}}(\sigma_i\frac{\partial\rho_{AB}(t)}{\partial t}\tau_j).
\end{eqnarray}
Where $\partial\rho_{AB}/\partial t$ has a similar expression with
Eq.(10) or Eq.(11), depending on what formalism you choose to
describe the time evolution of the system.

In summery, taking the decoherence effects into account, we study
dynamics of the entanglement rate for two special sets of initial
states. The interaction under consideration is of pairwise. The
results show that there are competitions between decohering and
entangling, those competitions lead to (1).For a specific
interaction and a decoherence scheme, some initial state could not
be used to prepare entanglement. (2). Some initial states can be
used to prepare or increase entanglement under a proper choice of
the parameters.

\end{multicols}

\end{document}